\documentstyle[12pt,psfig,aps]{revtex}
\begin{document}
%%\twocolumn

\title{A Quantum Full Adder for a \\Scalable Nuclear Spin Quantum Computer}

\author{{G.P. Berman$^1$, G.D. Doolen$^1$, G.V. L\'opez$^2$, and V.I. Tsifrinovich$^3$}}

\address{$^1$ T-13 and CNLS,  
Los Alamos National Laboratory, Los Alamos, New Mexico 87545}
\address{Departamento de F\'isica, Universidad de Guadalajara,
Corregidora 500, S.R. 44420, \\
Guadalajara, Jalisco, M\'exico}
\address{$^3$IDS Department, Polytechnic University,
Six Metrotech Center, Brooklyn NY 11201}

\maketitle

\begin{abstract}
\vspace{0.4cm}\\ 
We demonstrate a strategy for implementation a quantum full adder in a spin chain quantum computer. 
As an example, we simulate a quantum full adder in a chain containing 201 spins. Our simulations also demonstrate how one can minimize errors generated by non-resonant effects.
\renewcommand{\baselinestretch}{1.656} 


{keywords:~ qubit, full adder, quantum computer} 
\end{abstract}
\section{Introduction}
A full adder is a basic component of a conventional computer and a welcome asset for quantum computers. In particular, the Shor quantum algorithm requires modular exponentiation, $f(x)=a^x(\bmod N)$, which cannot be computed without a quantum full adder. The question arises: What is a full adder in a quantum computer? A quantum computer operates on a superposition of numbers simultaneously. What is not clear is what it means to add two superpositions. One definition of a full adder in a quantum computer is that a 
full adder is a gate which adds a given number to a superposition of numbers. This full adder must simultaneously add a definite number, ``a'', to all numbers, ``$b_i$'', which are in a superposition in a quantum computer register. The basic idea for the full adder is well-known \cite{1,2}. However it is not clear how to implement this idea in a many-qubit spin quantum computer. It is even less clear how to simulate accurately a quantum full adder having a large number of qubits using a conventional computer. One must understand the role of non-resonant effects and how to minimize them. One must also know the structure of the error states caused by non-resonant effects. Our paper provides an  answer for these issues. To implement a quantum full adder, we propose to use a sequence of electromagnetic $\pi$-pulses on a spin-chain quantum computer. We have simulated a dynamics of this quantum full adder with 201 spins, on a conventional computer; analyzed unwanted non-resonant effects; and determined the structure of the error states and ways to reduce non-resonant effects.

\section{A classical full adder}

A classical full adder operates with an input of two addend bits, ``a'' and ``b'', and a carry bit, ``c''. (See Table 1.)
In Table 1, $s^\prime$ and $c^\prime$ are the output sum and the carry-over, respectively. The sum, $s^\prime$, can be easily expressed as $s^\prime=a\oplus b\oplus c$ (where $\oplus$ is an addition modulo 2). The formula for a carry-over is more complicated. One can see from the Table 1, 
 that $c^\prime=ab\oplus ac\oplus bc$.

\section{A quantum full adder}

We describe the basic idea of a quantum full adder first suggested in \cite{1}. The full adder quantum gate (F) depends on the value of the bit ``a''. If $a=0$, the quantum computer applies the gate $F(0)$. If $a=1$, it applies $F(1)$. The F-gate can be expressed in terms of the Control-Not ($C_{ik}$) and Control-Control-Not ($C_{ikp}$) gates. A $C_{ik}$-gate changes the value of a target qubit, ``k'', if the control qubit, ``i'', has the value ``1''. A $C_{ikp}$-gate changes the state of the target qubit, ``p'', if both qubits, ``i'' and ``k'', have the value ``1'',
$$
C_{ik}|...n_i...n_k...\rangle=|...n_i...(n_i\oplus n_k)_k...\rangle,\eqno(1)
$$
$$
C_{ikp}|...n_i...n_k...n_p...\rangle=|...n_i...n_k...[(n_i n_k)\oplus n_p]_p...\rangle.
$$
\vspace{0.2cm}
$$
\begin{tabular}{|l|l|l||l|l|l|}\hline
a&b&c&$s^\prime$&$c^\prime$&$ab\oplus ac\oplus bc$\\ \hline
0&0&0&0&0&0\\ \hline
0&0&1&1&0&0\\ \hline
0&1&0&1&0&0\\ \hline
1&0&0&1&0&0\\ \hline
0&1&1&0&1&1\\ \hline
1&0&1&0&1&1\\ \hline
1&1&0&0&1&1\\ \hline
1&1&1&1&1&1\\ \hline
\end{tabular}
$$
\vspace{0.2cm}
\begin{center}
Table 1:~Table of binary addition.
\end{center}

We shall also use the Not-gate which can be designated by $C_i$. It changes the value of the $i$th qubit independent of the values of all other qubits. Using these gates, we can transform the state, $|0_3c_2b_1\rangle$, into the state, $|c^\prime_3s^\prime_2b_1\rangle$,
$$
F_{321}(a)\rightarrow |0_3c_2b_1\rangle=|c^\prime_3s^\prime_2b_1\rangle,\eqno(2a)
$$
where ``a'' and ``b'' are the addend bits and $s^\prime$ is their sum, $c$ and $c^\prime$ are the input and output carry bits. Namely,
$$
F_{321}(0)=C_{12}C_{123},~F_{321}(1)=C_{12}C_{123}C_2C_{23}.\eqno(2b)
$$
We use the convention that the right gate acts first. Let us check the action of the full adder,
$F_{321}$, for example, for $a=b=c=1$. We use $F_{321}(1)$:
$$
C_{23}|0_3c_2b_1\rangle=C_{23}|0_31_21_1\rangle=|1_31_21_1\rangle.\eqno(3)
$$
In (3), the second control qubit has the value ``1''. So, the third target qubit changes its value from ``0'' to ``1''. Next, 
$$
C_{2}|1_31_21_1\rangle=|1_30_21_1\rangle, C_{123}|1_30_21_1\rangle=|1_30_21_1\rangle,\eqno(4)
$$
$$
C_{12}|1_30_21_1\rangle=|1_31_21_1\rangle=|c_3^\prime s^\prime_2b_1\rangle.
$$
Thus, in accordance with the Table 1, we obtain,
$$
b=1,~s^\prime=1,~c^\prime=1.
$$

\section{A spin chain quantum computer}

An atomic chain quantum computer based on triplets, $ABC ABC ABC...$, was first introduced in \cite{3}. The implementation of this idea for a chain of spins interacting through the Ising interaction was given in \cite{4}. Ising-type spin chains have been used for quantum computation in a statistical ensemble quantum computers \cite{5,6}. In \cite{7}, we considered a simple model -- a chain of identical spins in a non-uniform magnetic field. 

The present paper is based on \cite{4} and \cite{7}. We consider a chain of spins (e.g. nuclear spins) in a non-uniform magnetic field. Similar to \cite{7}, the angle, $\Theta$, between the direction of the chain and the direction of the permanent magnetic field ($z$-direction) satisfies the condition: $\cos\Theta=1/\sqrt{3}$. Then, the dipole-dipole interaction between spins is suppressed, and the Ising interaction becomes dominant. We suppose that our chain consists of a periodic sequence of triplets, $ABC ABC ABC...$. The triplet, $ABC$, can be different nuclear spins or identical spins in slightly different environment. For definiteness, we shall keep in mind this second case \cite{7} with the following typical parameters,
$$
\omega_1/2\pi\sim 400MHz,~\omega_k=\omega_1+(k-1)\Delta\omega,~\Delta\omega/2\pi\approx 
20 kHz,\eqno(5)
$$
$$
J_{AC}/2\pi\approx 100Hz,~J_{BC}=2J_{AC},~J_{AB}=3J_{AC},
$$
where $\omega_1$ is the Larmor (angular) frequency of the right end spin, $\omega_k$ is the Larmor frequency of the $k$th spin, $J_{ik}$ is the constant of Ising interaction between the $i$th and $k$th spins. 

In the presence of a circularly polarized transverse magnetic field, the Hamiltonian of the system  can be written as \cite{8},
$$
{\cal H}=-\sum_k\omega_kI^z_k-2\sum_kJ_{k,k+1}I^z_kI^z_{k+1}-\eqno(6)
$$
$$
-{{\Omega}\over{2}}\sum_k\{I^-_k\exp[-i(\omega t+\varphi)]+I^+_k\exp[i(\omega t+\varphi)]\},
$$
where $I^z$ is the nuclear spin operator, $\omega$ and $\varphi$ are the frequency and the phase of the transverse magnetic field, $\Omega$ is the Rabi frequency (the magnitude of the transverse magnetic field in frequency units), $\hbar=1$.

\section{Implementation of the quantum full adder}

First we shall define our problem. Suppose that we have a number $|b\rangle=|b^{(L-1)}...b^{(0)}\rangle$, $b^{(m)}=0,1$. In decimal notation, $b=b^{(L-1)}2^{L-1}+...+b^{(0)}2^0$. (Below, we omit parentheses in the superscripts.) Note, that in general, we have a superposition of many numbers, $|b\rangle_i$, and any gate must act on all of these numbers simultaneously. We are going to add to a number $|b\rangle$, (all numbers, $b_i$, in a superposition) a definite number, $a=a^{L-1}...a^0$, where $a^m=0,1$. To achieve this goal we shall use $2L+1$ qubits. We load the number $b$ in a chain of qubits in the following way,
$$
|0_{2L+1}b_{2L}^{L-1}0_{2L-1}b^{L-2}_{2L-2}...0_5b_4^10_30_2b^0_1\rangle.
\eqno(7)
$$
This means that we place two additional qubits in the states $|0\rangle$ in front of the qubit $b^0_1$, and one additional qubit in the state $|0\rangle$ in front of all other qubits $b^m$, $m\not =0$. 

The gate $C_{123}$ is not convenient for our spin chain quantum computer in which each spin interacts only with its neighbors. So, we replace it with the following sequence of gates,
$$
C_{123}=C_{23}C_{32}C_{23}C_{132}C_{23}C_{32}C_{23}.\eqno(8)
$$
Let us check this equation, for example, for the state, $|0_31_21_1\rangle$. From the left side of Eq. (8) we have,
$$
C_{123}|0_31_21_1\rangle=|1_31_21_1\rangle,\eqno(9)
$$
as the first and the second qubits are in the states, $|1\rangle$, the third qubit changes its state. Next, we follow the operations on the right hand side of (8) to show that the same result is obtained.
From the right side of Eq. (8) we have,
$$
C_{23}|0_31_21_1\rangle=|1_31_21_1\rangle,~C_{32}|1_31_21_1\rangle=|1_30_21_1
\rangle,\eqno(10)
$$
$$
C_{23}|1_30_21_1\rangle=|1_30_21_1\rangle,~C_{132}|1_30_21_1\rangle=|1_31_21_1
\rangle,
$$
$$
C_{23}|1_31_21_1\rangle=|0_31_21_1\rangle,~
C_{32}|0_31_21_1\rangle=|0_31_21_1\rangle,
$$
$$
C_{23}|0_31_21_1\rangle=|1_31_21_1\rangle,
$$
which coincides with the right side of Eq. (9). Now, instead of (2b), we have the following expression for the full adder, $F$,
$$
F_{321}(0)=C_{12}C_{23}C_{32}C_{23}C_{132}C_{23}C_{32}C_{23},\eqno(11)
$$
$$
F_{321}(1)=C_{12}C_{23}C_{32}C_{23}C_{132}C_{23}C_{32}C_{23}C_2C_{23}.
$$

Now we explain how to add the numbers $b$ and $a$. If $a^0=0$, the quantum computer applies the gate, $F_{321}(0)$. If $a^0=1$, the quantum computer applies the gate, $F_{321}(1)$. According to (2), the result is,
$$
F_{321}|0_30_2b^0_1\rangle=|c^1_3s^0_2b^0_1\rangle,\eqno(12)
$$
where we have replaced $s^\prime$ by $s^0=a^0\oplus b^0$, and $c^\prime$ by $c^1$ which is the carry-over for the next addition of $b^1$ and $a^1$. 

Next, consider the five right most qubits,
$$
|0_5b^1_4c^1_3s^0_2b^0_1\rangle.\eqno(13)
$$
To add $b^1$ and $a^1$ and the carry-over, $c^1$, we should first make a swap, $S$, of the values of the fourth and the third qubits,
$$
S_{43}|b^1_4c^1_3\rangle=|c^1_4b^1_3\rangle.\eqno(14)
$$
The swap gate, $S_{ik}$, can be represented in terms of $C_{ik}$ gates,
$$
S_{ik}=C_{ki}C_{ik}C_{ki}.\eqno(15)
$$
Let us check, for example, the action of the $S_{21}$ gate on the state, $|0_21_1\rangle$. Using (15), we have,
$$
S_{21}=C_{12}C_{21}C_{12},\eqno(16)
$$
$$
C_{12}|0_21_1\rangle=|1_21_1\rangle,~C_{21}|1_21_1\rangle=|1_20_1\rangle,~
C_{12}|1_20_1\rangle=|1_20_1\rangle.
$$
Thus, the $S$-gate transforms the state, $|01\rangle$, into the state, $|10\rangle$. 

After the action of the $S_{43}$ gate, the state (13) changes into,
$$
|0_5c^1_4b^1_3s^0_2b^0_1\rangle.\eqno(17)
$$
The state, $|0_5c^1_4b^1_3\rangle$, has the form (2a), and it is ready for application of the full adder, $F_{543}$,
$$
F_{543}|0_5c^1_4b^1_3\rangle=|c^2_5s^1_4b^1_3\rangle.\eqno(18)
$$
Certainly, we use $F_{543}(0)$ if $a^1=0$ and $F_{543}(1)$ if $a^1=1$. Thus, we obtain the sum, $s^1=a^1\oplus b^1\oplus c^1$, and a carry-over, $c^2$, for the next addition. It is clear that by repeating the application of gates $F$ and $S$, we shall obtain the desired answer. A complete full adder, $F$, can be represented as a combination of the elementary full adders, $F_{ijk}$, and the swap-gates, $S_{ij}$,
$$
F=F_{2L+1,2L,2L-1}(a^{L-1})S_{2L,2L-1}...\eqno(19)
$$
$$
F_{765}(a^2)S_{65}F_{543}(a^1)S_{43}F_{321}(a^0).
$$
After the action of the full adder, the superposition of the states (7) transforms into the superposition of states,
$$
|c^Ls^{L-1}b^{L-1}s^{L-2}b^{L-2}...s^0b^0\rangle.\eqno(20)
$$
Thus, the qubits in even positions and the left end qubit carry the results of addition.
The problem is: How to implement all of these gates using $\pi$-pulses? We suppose that $\Omega\ll J_{ik}$, say $\Omega/2\pi\approx 10$Hz. This allows us to excite each spin individually. To implement $F_{321}$, we use equations (11). As an example, we show how to implement  $F_{321}(0)$. According to Eqs (11), first we implement $C_{23}$, then $C_{32}$, and so on. Suppose we have an integer number of triples, $ABC$, in our spin chain. Then, we have 4 possible frequencies in the position $(3n-2)$ which correspond to a spin $C$:
$$
\omega^{00}_{3n-2}=\omega_{3n-2}+J_{BC}+J_{AC},\eqno(21)
$$
$$
\omega^{01}_{3n-2}=\omega_{3n-2}+J_{BC}-J_{AC},
$$
$$
\omega^{10}_{3n-2}=\omega_{3n-2}-J_{BC}+J_{AC},
$$
$$
\omega^{11}_{3n-2}=\omega_{3n-2}-J_{BC}-J_{AC}.
$$
Here $\omega^{ij}_k$ corresponds to the $k$th spin whose left neighbor is in state $|i\rangle$ and whose the right neighbor is in state $|j\rangle$. Similar expressions can be found for spins in positions $(3n-1)$ and $(3n)$, the $B$ and $A$ spins, respectively. The end spins have only two frequencies. For the left end spin, $A$,
$$
\omega_{2L+1}^0=\omega_{2L+1}+J_{AB},~\omega_{2L+1}^1=\omega_{2L+1}-J_{AB},\eqno(22)
$$
and for the right end spin, $C$,
$$
\omega_{1}^0=\omega_{1}+J_{BC},~\omega_{1}^1=\omega_{1}-J_{BC}.\eqno(23)
$$

Now, to implement $C_{23}$, we apply a $\pi$-pulse with frequency $\omega^{01}_3$ and then a $\pi$-pulse with the frequency $\omega^{11}_3$. One of these two pulses definitely changes the state of the third spin if the second spin is in its excited state, $|1\rangle$. To implement the gate $C_{123}$, we need a single $\pi$-pulse with frequency $\omega^{11}_2$. To implement a Not-gate, $C_2$, which appears in $F_{132}(1)$, we have to apply four $\pi$-pulses with all possible frequencies, $\omega^{ij}_2$, $i=0,1$, and $j=0,1$. The total number of pulses required to implement a $F_{ijk}(0)$-gate (if $i\not=2L+1$, $k\not=1$) is 15: two pulses for each Control-Not gate and one pulse for the Control-Control-Not gate. For the $F_{ijk}(1)$ gate, the total number of pulses is 21 (16 pulses for the left triple of the chain). The swap gate, $S_{ij}$, requires 6 pulses.

Thus, to add $L$-qubit numbers, our proposal requires $(2L+1)$ qubits and less than $27L$ $\pi$-pulses. 

\section{Simulations of a quantum full adder}

For numerical simulations of a quantum full adder we used the following assumption: the frequency difference between two neighboring spins is much greater than the Rabi frequency. As a result, the selective excitation of a chosen spin has a small effect on all other spins. At the same time, we take into account the action of a $\pi$-pulse on non-resonant states. For example, suppose that the frequency of a $\pi$-pulse is: $\omega=\omega^{11}_3$. This pulse is resonant with spin ``3'' only if the neighboring spins, ``1'' and ``2'', are in their excited states. The states in which the spin ``3'' has frequencies, $\omega^{10}_3$, $\omega^{01}_3$,
and $\omega^{00}_3$, are non-resonant for the pulse with $\omega=\omega^{11}_3$. We take into consideration the transformations of all these non-resonant states.

Note, that any pulse in our simulations acts on all basic states in the quantum superposition.  The resonant state transforms into a state with the opposite direction of the resonant spin. Every non-resonant state transforms into the superposition of two states: the initial one and an error state generated by the non-resonant transition. If the probability of an error state is less than a chosen small number, $\varepsilon$, our computer program automatically removes it. One can argue that every removed state can generate a number of new states. 
However, suppose that we have removed a state with a small probability, $P$. Due
to the main property of the unitary transformations, the sum of
the probabilities of all states generated by the removed state (including
itself) is $P$. Thus, the total probability of all neglected states cannot
increase in spite of generation of new states!

We have simulated the addition of the 100-digit numbers in a 201 spin chain. To minimize non-resonant errors, we used the $2\pi k$-method \cite{7,8}. The basic idea of the $2\pi k$-method is the following. One chooses the Rabi frequency, $\Omega$, of a resonant $\pi$-pulse in such a way that it becomes approximately a $2\pi k$-pulse for all non-resonant transitions (where $k$ is an integer which generally is different for different states). A $2\pi k$-pulse does not generate unwanted error states. A transformation of a basic state under the action of a $\pi$-pulse was described in our previous paper \cite{7}, where the following values of dimensionless parameters were chosen,
$$
J_{AC}=1,~J_{BC}=2,~J_{AB}=3.\eqno(24)
$$
We found that all non-resonant transitions approximately satisfied
the $2\pi k$-conditions for the value of the Rabi frequency, $\Omega=\Omega_0=0.10005$. 

\begin{figure}
\psfig{file=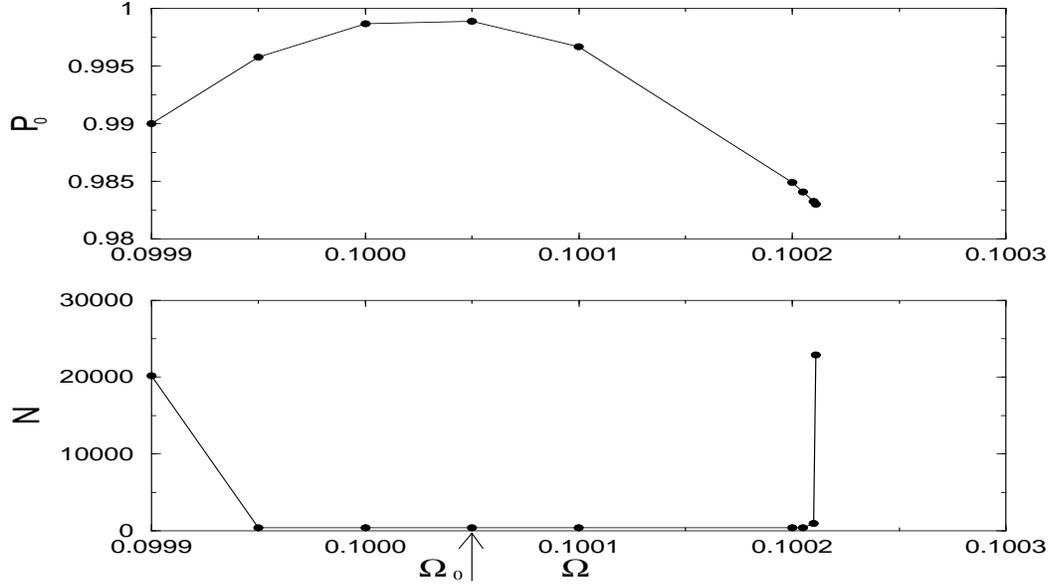,width=14cm,height=8cm}
\caption{The probability of the expected state, $P_0$; and the number of error states, $N$, as a function of the Rabi frequency, $\Omega$.} 
\label{Fig:1}
\end{figure}

Next, we present the results of our numerical simulations. As an example, we add the ``classical number'',
$$
1^{99}0^{98}...0^0,\eqno(25)
$$
($2^{99}$ in a decimal notation) to the ``quantum number'',
$$
0^{99}...0^11^0,\eqno(26)
$$
($1$ in a decimal notation). The sum of these two numbers,
$$
1^{99}0^{98}...0^11^0,\eqno(27)
$$
corresponds to the quantum state of the chain of 201 spins,
$$
|0_{201}1_{200}0_{199}...0_31_21_1\rangle.\eqno(28)
$$
\begin{figure}
\psfig{file=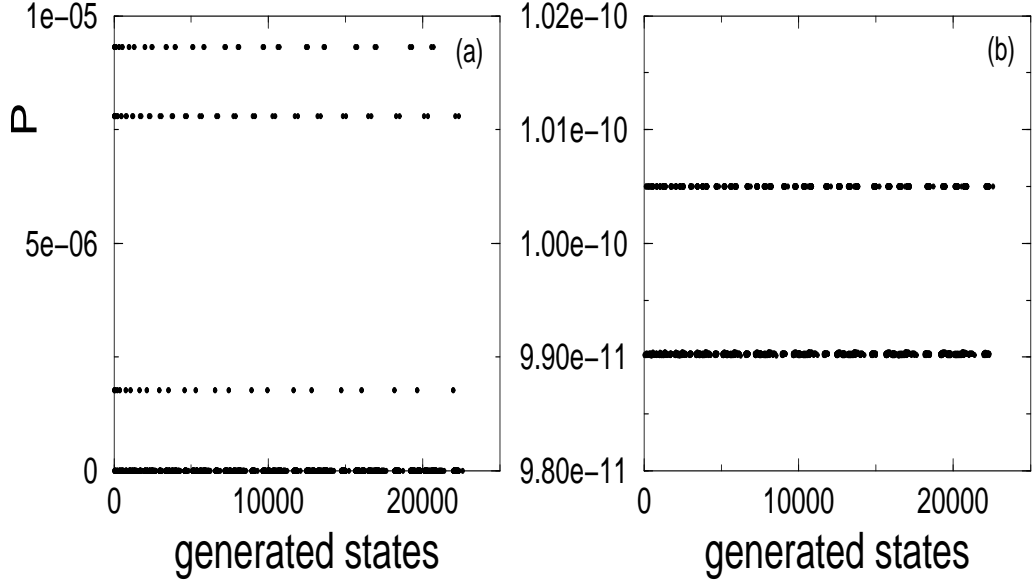,width=14cm,height=8cm}
\caption{``Line spectrum'' of the probabilities, $P$, of error states $(\Omega=0.10021$); (a) the region: $P\sim 10^{-6}$, (b) the region: $P\sim 10^{-10}$.} 
\label{Fig:2}
\end{figure}

 For the value of $\Omega=\Omega_0$ which approximately satisfies $2\pi k$-condition, we have for the probability of the expected state (28): $P_0=0.99889$. The number, $N$, of error states with probability $P>10^{-12}$ is only 304. Small deviations from the $2\pi k$-condition significantly influence the result. As an example, for $\Omega=0.10021$, the value of $P_0$ decreases to $0.98300$, and the number of error states, $N$, grows to $46530$. 
Fig. 1 shows the dependence of $P_0$ and $N$ on the Rabi frequency, $\Omega$ in the vicinity of $\Omega_0$. One can see that the probability, $P_0$, smoothly decreases by approximately $1\%$ when the Rabi frequency, $\Omega$, shifts from $\Omega_0$ by approximately $0.1\%$. Unlike the value of $P_0$, the number of error states with probability $> 10^{-12}$ does not change in the close vicinity of $\Omega_0$. But it  sharply increases when deviation of $\Omega$ from $\Omega_0$  approaches $0.15\%$. Fig. 2 shows the probabilities of error states (the states are shown in the order of their generation). One can see a specific ``line spectrum'' of the probabilities: the error states are ``attracted'' to a few discrete values of the probability. Very similar effects were obtained for other examples of addition including the addition of a ``classical'' number to a superposition of two quantum numbers.

\section{Phase control}

The quantum full adder implementation considered above provides a transformation of an arbitrary superposition of ``quantum numbers'', $q_k$, into a superposition of numbers $(q_k+a)$, where $a$ is any given ``classical number''. However, this proposed scheme also generates complicated phase differences between the states of the quantum superposition. This effect can be inappropriate, especially for the Shor algorithm. In this section, we describe how to extend our simple model to incorporate phase restoration after the action of every $\pi$-pulse.

Consider the chain of paramagnetic ions ABCABC containing nuclear spins
in a high nonuniform magnetic field. Every electron spin interacts with
the nuclear spin of its ion via the hyperfine interaction. Electron
spins interact with each other through the Ising interaction. All electron
spins are in their ground states. The nuclear spins also interact with each other
through the Ising interaction. This interaction is responsible for
quantum logic gates. The dipole-dipole interaction is suppressed due to the
use of the magic angle between the chain and the direction of the external magnetic
field. The key point of our model is the following: we consider the interaction
(for simplicity, the Ising interaction) between an electron spin and its
two neighboring nuclear spins. This interaction can originate, for
example, if the electron density at the neighboring nuclei is not zero. Thus, the electron spin frequency for a particular ion can take eight values. As an example, for the B-ion one has, 
$$
\omega^{000}=\omega_e+J^{AB}_{ee}+J^{BC}_{ee}+A^B/2+J^{AB}_{en}+J^{BC}_{en},\eqno(29)
$$
$$
\omega^{100}=\omega_e+J^{AB}_{ee}+J^{BC}_{ee}-A^B/2+J^{AB}_{en}+J^{BC}_{en},
$$
$$
\omega^{110}=\omega_e+J^{AB}_{ee}+J^{BC}_{ee}-A^B/2-J^{AB}_{en}-J^{BC}_{en},
$$
and so on. Here $\omega^{ijk}$ corresponds to the electron frequency for the case in which the nuclear spin of the same ion is in the state $|i\rangle$, the nuclear spin of the neighboring $A$-ion is in the state $|k\rangle$, and the nuclear spin of the neighboring $C$-ion is in the state $|j\rangle$; $\omega_e$ is the Zeeman frequency, $J^{AB}_{ee}$ and $J^{BC}_{ee}$  are the constants of the electron-electron interaction, $A^B$ is the hyperfine constant (in the frequency units),  and $J^{AB}_{en}$ and $J^{BC}_{en}$  are the constants of the electron-nuclear interaction for neighboring ions.

Thus, the ESR frequency depends on the position of an electron spin in the
chain (because the Zeeman frequency is nonuniform) and the states of three
nuclear spins -- the nuclear spin of its own ion (via the hyperfine
interaction) and the nuclear spins of two neighboring ions. One can tune
the frequency of an electromagnetic pulse in such a way that it is
resonant with the electron spin only if it is in a definite position in
the chain and the three nuclear spins mentioned above are in the definite
states.

The strategy for phase correction for the quantum full adder is the following.
The full adder is implemented by a sequence of the nuclear $\pi$-pulses. A
nuclear $\pi$-pulse causes a phase shift of $\pi/2$ for the resonant states. There are six possible
phase distortions for non-resonant states (3 possible states for
neighboring nuclear spins times two possible
states for the selected nuclear spin). Correspondingly, one has six possible
frequencies for the electron spin of the selected ion in the non-resonant
state. Because of the large value of the electron gyromagnetic ratio, we
assume that each of the corresponding electron transitions can be
driven without noticeable non-resonant effects on the electron
transitions with close frequencies. After the nuclear $\pi$-pulse, one
applies 12 electron $\pi$-pulses: two $\pi$-pulses for every possible
frequency of the selected electron spin in the non-resonant state. The
phase of the first $\pi$-pulse is zero, the phase of the second one is $\phi$.
The total phase shift of the wave function of the ion is $(\pi + \phi)$. By choosing
an appropriate value for $\phi$, one can change the phase shift for a
specific non-resonant state to the value of $\pi/2$, which corresponds to
the resonant state. After the action of 12 electron $\pi$-pulses,  all states in the quantum superposition will have the same phase, and all electron
spins will be in their ground state. Thus, the phase distortion generated by a nuclear $\pi$-pulse can be corrected with 12 electron $\pi$-pulses.

\section{Conclusions}

We have demonstrated a strategy for implementation of a full adder in a nuclear spin quantum computer. To add an arbitrary superposition of L-qubit ``quantum numbers'' and a fixed L-bit ``classical number'' our scheme requires 2L+1 qubits and less than 27L resonant $\pi$-pulses.
We have simulated the action of this full adder for a spin chain containing 201 qubits. Using the $2\pi k$-method allowed us to minimize the generation of unwanted non-resonant error states. These error states are shown to have a few preferred discrete values of probability (a ``line spectrum''). Our simple spin model is generalized to correct undesired phase distortions.
\section*{Acknowledgments}
The work  was supported by the Department of Energy (DOE) under contract W-7405-ENG-36, by the National Security Agency (NSA) and by the Advanced Research and Development Activity (ARDA).
\end{document}